\documentclass[format=sigconf,anonymous=false,review=false,nonacm]{acmart}
\usepackage{hyperref} 

\AtBeginDocument{%
  \providecommand\BibTeX{{%
    \normalfont B\kern-0.5em{\scshape i\kern-0.25em b}\kern-0.8em\TeX}}}

\begin{document}

\title{Looking for Complexity at Phase Boundaries in Continuous Cellular Automata}


\author{Vassilis Papadopoulos}
\affiliation{%
  \institution{Ecole Polytechnique Fédérale de Lausanne (EPFL)}
\country{Switzerland}}
\orcid{0000-0003-3270-5219}
\email{vassilis.papadopoulos@epfl.ch}

\author{Guilhem Doat}
\affiliation{%
  \institution{Ecole Polytechnique Fédérale de Lausanne (EPFL)}
  \country{Switzerland}
}
\email{guilhemdoat@gmail.com}

\author{Arthur Renard}
\affiliation{%
  \institution{Ecole Polytechnique Fédérale de Lausanne (EPFL)}
  \country{Switzerland}
}
\email{arthur.renard@epfl.ch}

\author{Cl\'ement Hongler}
\affiliation{%
  \institution{Ecole Polytechnique Fédérale de Lausanne (EPFL)}
\country{Switzerland}}
\email{clement.hongler@epfl.ch}







\renewcommand{\shortauthors}{Papadopoulos, Doat, Renard, Hongler}
\newcommand{\guil}[1]{\textcolor{red}{#1}}
\newcommand{\arth}[1]{\textcolor{blue}{#1}}
\begin{abstract}
One key challenge in Artificial Life is designing systems that display an emergence of complex behaviors. Many such systems depend on a high-dimensional parameter space, only a small subset of which displays interesting dynamics. Focusing on the case of continuous systems, we introduce the 'Phase Transition Finder'(PTF) algorithm, which can be used to efficiently generate parameters lying at the border between two phases. We argue that such points are more likely to display complex behaviors, and confirm this by applying PTF to Lenia showing it can increase the frequency of interesting behaviors more than two-fold, while remaining efficient enough for large-scale searches.
\end{abstract}


\keywords{Lenia, Artificial, Life, Alife, Complex Systems, Phases}

\begin{teaserfigure}
\centering
  \includegraphics[width=0.95\textwidth]{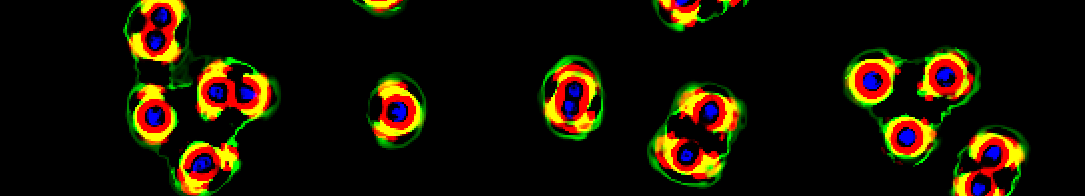}
  \caption{ Cell-like creatures in Lenia, displaying behaviour reminiscing of mitosis}
  \Description{Lenia}
  \label{fig:teaser}
\end{teaserfigure}


\maketitle

\section{Introduction}
Among the main goals of the field of Artificial Life (Alife) lies the discovery, or design, of systems that display 'interesting' life-like behaviors, such as the spontaneous emergence of creatures that interact in complex manners \cite{gardner_mathematical_1970}. To this end, a plethora of Alife systems were carefully designed to exhibit some kind of emergent property; be it Turing completeness \cite{neumann:theory,wolfram_universality_1984}, self-organization \cite{mordvintsev_growing_2020}, self-reproduction \cite{langton_self-reproduction_1984} or simply the appearance of solitons \cite{chan_lenia_2020}. Despite those impressive results, manually designing such systems is very difficult, as it is often hard to predict what parameters will result in interesting emerging behaviors.

To address this problem, some more automatized approaches are being developed, leveraging the increase in computing power, as well as the recent advances in machine learning. For instance, genetic algorithms are effective to efficiently traverse the parameter space of an Alife system, when we are looking for a specific behaviour\cite{hamon_learning_2022}. Such methods can also be combined with 'novelty maximization' techniques\cite{yang_self-replicating_2023}, which presumably allow exploring the parameter space 'efficiently', visiting only novel behaviors. One can also employ gradient-descent based methods to steer the dynamics towards a desired goal \cite{mordvintsev_growing_2020}, or combine it with a more abstract target to again obtain an efficient 'parameter-exploring' algorithm \cite{etcheverry_hierarchically_2020}.

While such methods produce 'interesting' systems much more reliably than random sampling, they remain relatively slow; usually, the generation of a new set of parameters requires a new training run, which is slow (especially for genetic algorithms), and often expensive to scale. In this paper, we explore a complementary method to generate interesting systems at a much higher rate than random sampling, while remaining scalable and lightweight as it does not rely on machine learning methods.

The main idea we seek to exploit is the observation that \emph{phase transition regions} often display complex dynamics. One potential reason for this peculiar observation comes from statistical physics, from which we learn that (infinite) systems at a phase transition point display scale invariance\cite{landau_theory_1937}. In particular, this implies the existence of structures at every scale, which is certainly non-trivial emergent behaviour if the dynamics are local. Ideas from self-organized criticality \cite{bak_self-organized_1989} also seem to suggest that critical points are an important driver of complexity. In this paper, we design a 'proof-of-concept' algorithm for homing in these regions in parameter space and confirm empirically that the proportion of 'interesting' behaviors is much higher than random sampling.

The structure of this paper is as follows. In section 2, we describe the 'Phase Transition Finder' (PTF) algorithm in more detail and develop a concrete implementation for parameter exploration in Lenia. In section 3, we present the results of the application of PTF on Lenia, and assess its efficacy and scalability. We conclude by discussing potential applications and future research directions.

\section{Idea and algorithm}
It is a well-known empirical fact that the 'interesting' dynamics in Alife systems are much rarer than their chaotic or periodic counterparts. For instance, in the class of outer-holistic 2D cellular automata estimations put the number of interesting dynamics at $\sim1\%$\cite{hudcova_classification_2020}. In continuous systems such as Lenia\cite{chan_lenia_2019}, mapping of a slice of parameter space shows that creatures appear only in a very thin part of the full region. Moreover, whenever a set of parameters that displays emergence of creatures is discovered, they are usually quite fragile; a slight change in parameters can make the creature suddenly die, or expand uncontrollably. Such fine-tuning is again reminiscent of a critical point between two phases, where very slight changes can topple the system in one phase or the other.

\subsection{Phase Transition Finder Algorithm}
These observations lead us to conjecture that at the very least, points lying at the edge between two phases are worth exploring for interesting dynamics. We seek to design an effective method of homing into these regions which we will call 'Phase Transition Finder' (PTF). The first step in this direction is of course determining two (or more) mutually exclusive \emph{phases}. We choose a definition of 'phase' that is intentionally very loose such that these ideas are applicable to a wide range of models. We define it simply as a quantitative description of some aspects of the dynamics of the model. To give a few examples, we could say we are in a 'dead' phase if all activations of a cellular automaton go to $0$ after a certain time. Conversely, we could define an 'exploding' phase, if more than $80\%$ of cells are alive after a certain time. For multi-channel models, we could define different phases according to which channel is dominant, according to the growth rate in the initial stages, or the magnitude of the fluctuations in the final stages... The possibilities are endless.

\begin{figure}[h!]
  \centering
  \includegraphics[width=0.6\linewidth]{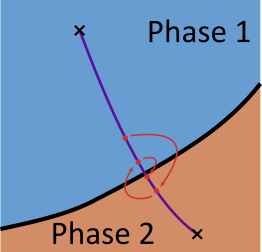}
  \caption{2-dimensional parameter space for an Alife model. Two points in different phases are located, and then by dichotomy, we rapidly approach the phase transition region.}
  \Description{Figure showing a path in 2-dimensional alife model parameters space.}
  \label{fig:dichotomy}
\vspace{-0.4cm}
\end{figure}

The second step is to sample points from the transition regions of the phases we have defined. This is relatively straightforward; we begin by randomly sampling points from a pre-determined region of parameter space (the prior region), until we find two points that lie in different phases. Next, we connect them by an arbitrary path (in parameter space), and by dichotomy, we efficiently find the point lying at the interface between the two phases (see Fig.\ref{fig:dichotomy})\footnote{Note that here we could also use gradient descent. However, that might not always be applicable as the definition of the phases (which is binary in nature) does not always provide with a differentiable proxy.}. Crucially, this step relies on the \emph{continuity} of the parameter space\footnote{Still, it would be interesting to apply it on discrete parameter landscapes, as it may provide interesting results, although the 'transition' point is not guaranteed to exist.}. For this method to work, we need to choose a set of mutually exclusive phases that cover the full parameter space. Additionally, if the algorithm is to be efficient, determining to what phase a point belongs should be fast. For this, it is okay to use heuristic methods, as having perfect accuracy is not necessary.

Note that the choice of the prior region is very important; ideally one should pick a prior region that displays an equal amount of each of the chosen phases. This is a weakness of this method, as the choice of the prior must be done manually. This can be mitigated whenever the chosen phases are sufficiently simple such that one can roughly guess their locations and extent. In the discussion at the end we mention some directions that could be useful in solving this weakness.

\subsection{An illustrative example: multi-channel Lenia}
To illustrate PTF and verify its usefulness, we design a proof-of-concept experiment by focusing on the class of continuous Alife systems known as (multi-channel) Lenia. Lenia was developed by Bert Chan \cite{chan_lenia_2020}, first as a natural continuous extension of the Game of Life. This model displays a wide variety of emerging creatures and gliders, and was later extended from a single channel to multiple ones, increasing further the diversity of its 'fauna'. We remind here how multi-channel Lenia works, as it will be the focus of our experiments.

The state of the world at time $t$ is fully determined by a 3-channel field where each channel is represented by $A^t_i(\vec{x})$ taking values in $[0,1]$, where $i$ indexes the three (R,G,B) channels, and $\vec{x}=(x,y)$ is the coordinates of the 2D world. The state at $t+dt$ is computed according to the following equation :
\begin{equation}
    A^{t+dt}_i(\vec{x}) = \bigg[ A^{t}_i(\vec{x})+ dt \sum_j B_{ji} G_{ji}( K_{ji} \ast A^t_j)\bigg]^1_0
\label{eq:LeniaMultiDef}
\end{equation}

\begin{figure}[h!]
  \centering
  \includegraphics[width=0.8\linewidth]{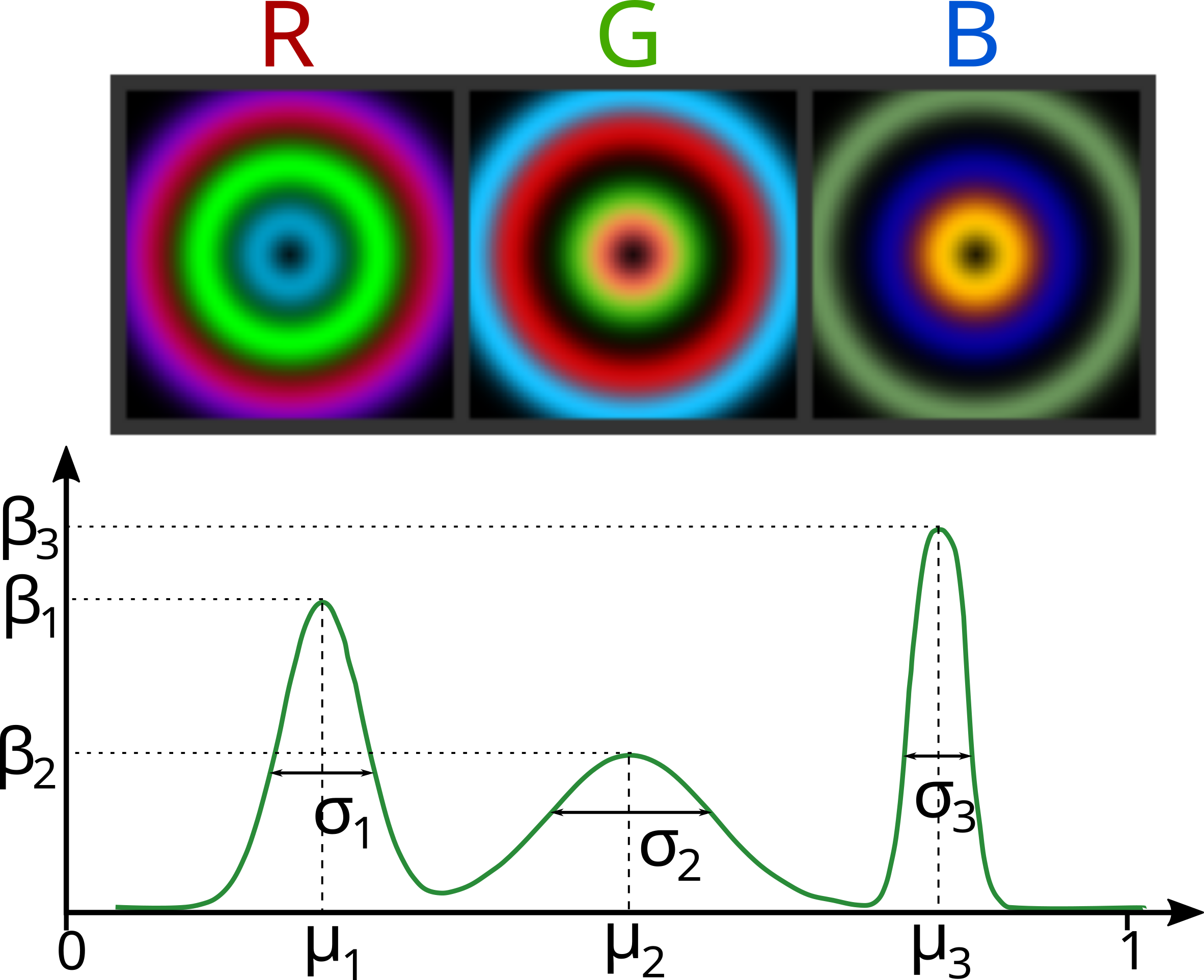}
  \caption{\textbf{Top}: Example kernels $K_{ji}$. Each kernel is convoluted with the corresponding labeled channel on top. The colors of the circles indicates to which channel the convolution contributes (i.e., the '$i$' in $K_{ji}$). \textbf{Bottom} : Cross-section of one kernel $K_{ji}$, displaying the 8 free parameters ($\sum_j \beta_j=1$).}
  \Description{Figure depicting a Lenia Kernel and how it is constructed.}
  \label{fig:kernels}
  \vspace{-0.4cm}
\end{figure}

Where $K_{ji}(\vec{x}) \in \mathbb{R}^2\rightarrow \mathbb{R}$ are kernels that compute the contribution of channel $j$ for channel $i$ through a convolution. They can be arbitrary radially symmetric functions, but to reduce the parameter space dimensionality, they are entirely determined by the position and magnitudes of three bump functions $\exp{-\frac{(x-\mu)^2}{2\sigma^2}}$, see Fig.\ref{fig:kernels}. Similarly, $G_{ji} =2\exp{\frac{(x-\mu_{ji})^2}{2\sigma_{ji}}}-1 \in [-1,1]$ are growth functions, which are there to introduce non-linearities in the dynamics. Again, they are fully determined by specifying $\mu_{ij}$ and $\sigma_{ij}$. The $B_{ij} \in [0,1] \mbox{s.t.} \sum_j B_{ij} = 1$ weight the contributions of the different channels. Finally, $\left[\right]^1_0$ denotes clipping to $[0,1]$ such that the values of $A$ remain in the same range (which also introduces an additional non-linearity). 

In total (excluding the choice of $dt$, which we will keep fixed), this amounts to $106$ real numbers in $[0,1]$ as free parameters. Navigating such a high-dimensional manifold manually in search of interesting dynamics is not an easy task. Luckily, this particular model suggests natural phases for the phase-transition search, those being 'dead' and 'living' phases. This is because moving creatures in Lenia exist necessary in the balance between death and expansion, as they destroy 'matter' behind them and recreate it in front, simulating locomotion. 

More precisely, we formally define the 'dead' phase as $\lim_{t\rightarrow \infty} \bar{A}^t_i < \alpha$, with $\alpha$ a hyperparameter and $\bar{A}^t_i$ denotes the average over the spatial positions. The 'live' phase is defined as the complement of the 'dead' phase. In practice, to determine efficiently which phase we are in, we approximate the definition by cutting at finite time $T$, since it is otherwise technically undecidable.

When we attempt to evaluate by inspection dynamics that we find using this technique, we will be looking for the following qualities, which aim to capture why some dynamics look interesting:
\begin{itemize}
    \item Dynamics that do not settle quickly
    \item Several coexisting and interacting colors
    \item Presence of solitons, gliders
    \item Self-reproduction and other interesting interactions between solitons
\end{itemize}

\section{Results on Lenia}
In this section, we present some results stemming from the application of the algorithm to multi-channel Lenia. For these experiments, we decided to explore only the space of parameters affecting the \emph{physics} of the system, namely the parameters appearing in \ref{eq:LeniaMultiDef}. One could extend the search to consider the initial condition $A^0_i(\vec{x})$, which can also be impactful, but we wanted to focus on finding interesting 'physics', rather than interesting initial conditions. Thus, we will be initializing all runs with Perlin noise in all channels. In this way, we are implicitly selecting for interesting dynamics that develop for this class of random initial conditions. 

We implement multi-channel Lenia using Pytorch (code provided in supplemental material). For the discretization, we set the scale by choosing the kernel to be of size $1\times 1$ (from here on, unless specified, all units are in 'kernel size'). We fix $dt=0.1$ and discretize the kernel to a $31\times31$ pixels square, equivalent to $dx \approx 0.03$. We perform the visual evaluation on $15$ second clips ($1800$ timesteps at $120$ fps), on a $\approx 16\times 16$ toroidal world. 

To speed up the search for the alive and dead phases, we restrict to a $5\times 5$ world, evolved for 800 timesteps. This can be justified because the scale of emergent structures is rarely much larger than a few kernel sizes, so restricting to a smaller world does not affect the outcome for which phase we are in. In any case, a few misclassifications are unproblematic, since we do not aim for $100\%$ success rate in the parameters we discover.

We begin with a relatively 'neutral' prior. We set all parameters to be generated uniformly in $[0,1]$, except $\mu$ and $\sigma$ for the growth functions, which are uniform in $[0,0.7]$, $[0,0.2]$ respectively (specifics in the provided code), otherwise 'dead phases' are extremely rare. We set the threshold to $\alpha=0.05$ and generate 200 phase transition points. We then qualitatively assess the amount of 'interesting' dynamics for these parameters against 200 random parameters. 

Next, we repeat the experiment with a slightly 'smarter' prior. One can see that if any of the $G_{ij}(0)>0$, then the 'ground state' is unstable; there can be no 'dead' phase, as a dead world will have a non-negative growth! The critical value in this case is given by $\sigma_{ij} = \frac{\mu_{ij}}{2\ln(2)}$, we thus constrain the prior for $\sigma_{ij} \in [0,0.9\frac{\mu_{ij}}{2\ln(2)}]$. This choice ensures both the a phase transition is present in the prior, and that the proportion of dead/alive phases will be roughly balanced, speeding the initial phase search. We repeat the experiment as before displaying results in Table \ref{tab:results}.

\begin{figure}[h!]
  \centering
  \includegraphics[width=0.8\linewidth]{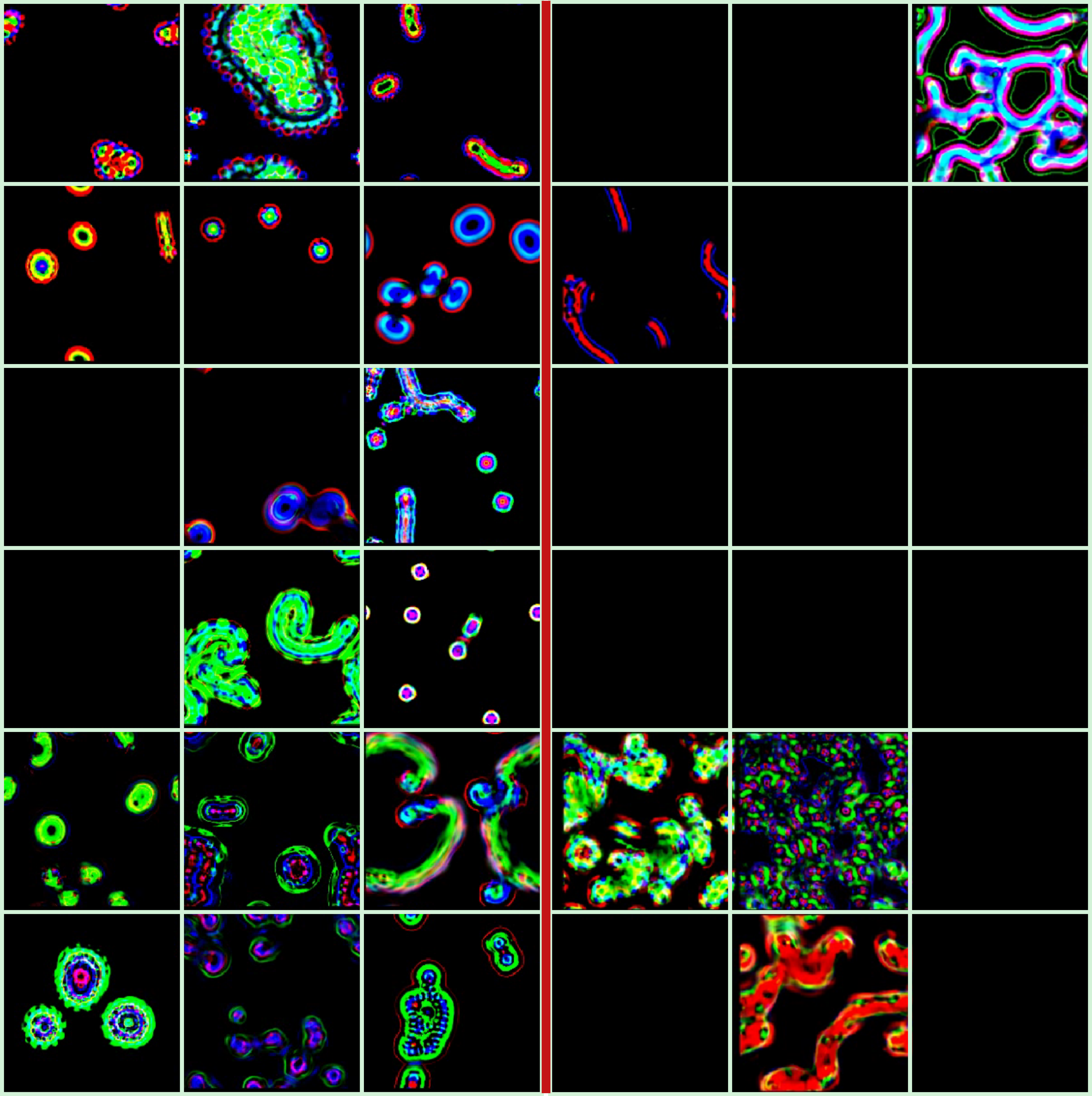}
  \caption{Final snapshot for a simulation of 1000 steps in Lenia. Left of the red line: parameters obtained with PTF. Right : Parameters obtained with prior.}
  \Description{Figure showing a path in 2-dimensional alife model parameters space.}
  \vspace{-0.5cm}
  \label{final_state}
\end{figure}

\begin{table}[h!]
\begin{tabular}{|l|l|l|l|l|}
\hline
Prior   & \% Dead & \% Int. rand & \% Int. phase & Time (m:s) \\ \hline
neutral &     32\%    &    5.5\%         &    13.0\%          &  4:40    \\ \hline
smart   &     65\%    &    14.0\%         &     30.5\%         &  4:20    \\ \hline
\end{tabular}
\caption{ Table with results for the experiments. "$\%$ Dead" denotes the proportion of "dead" phase when sampling from the prior. "$\%$ Int. rand" (resp. phase) denotes the proportion of dynamics deemed 'interesting' by visual inspection, when sampling from the prior (resp. from phase transition points). Time columne is the time taken to generate 100 phase transition points on a RTX 3080 GPU.}
\label{tab:results}
\vspace{-0.7cm}
\end{table}
Overall, we can see that the points in the phase transition region display a significantly higher percentage of 'interesting' configurations (see Figure \ref{final_state} for a glimpse). Additionally to these controlled tests, we extensively tested the algorithm for different priors. Overall, qualitatively we observe that for phase transition points, the presence of solitons, static or even moving is not uncommon, this is in stark contrast to a random parameter search, where finding gliders is extremely uncommon. Some videos of remarkable dynamics that were found using this method can be found in the supplementary material.
One thing we noticed is that the phase transitions are often not 'sharp', but rather continuous. This is to be expected as nothing guarantees our choice of phase to match in any way the 'rigorous' definition in statistical physics or thermodynamics, especially for arbitrary $\alpha>0$. When $\alpha$ is close to zero, we sometimes see behavior reminiscent of 'critical slowing down': close to the transition point, the dynamics slow down and die out (or conversely, expand) slowly, such that at the cutoff time $T$, we are indeed at $\lim_{t\rightarrow \infty} \bar{A}^t_i \approx \alpha$. This is not what we are looking for as it generally does not give rise to interesting dynamics, but there is no way to discard those points a priori. 

In contrast, when the parameters display solitons, the phase transition is usually sharp : small fluctuations in parameter space usually kill off the solitons, or make them unstable to nucleation. One last remark is that self-reproduction behaviour (usually through duplication of the solitons) is quite a common occurence. It is unclear if this is a specificity of Lenia or of the phase-transition regions, and it would be interesting to test this in other continuous models.

\section{Discussion and research directions}
In this short paper, we presented a novel method for the automatic search of interesting sets of parameters in continuous Alife systems. We presented some encouraging results by applying it to Lenia, validating the potential of this method. In particular, the rate at which new points in parameter space can be constructed is relatively high ($\sim$40 minutes for 1000 examples, although this can highly vary depending on the hyperparameters we choose), which opens many avenues to apply deep learning techniques. Indeed, the main problem with pure 'random' generation of parameters is that the 'signal' in them is small; generally, the amount of interesting dynamics is extremely low. In this case, using an efficient method like the one presented in this paper can be useful for increasing the 'signal' in the dataset to a more tractable level for machine learning applications.

The algorithm does however have some limitations, the first of which is that the choice of the prior region still has an important impact on the results, and it still requires some degree of manual tuning. This can potentially be mitigated by developing heuristics for the choice of the prior region, such as modifying it until the two phases cover approximately the same volume. Another problem is that a large part of the generated parameters still display trivial dynamics. Some (like dying dynamics) can be easily detected and discarded. For other less obvious cases, one possible direction to pursue is to train a ML model to discard uninteresting dynamics, by learning through human feedback\cite{christiano_deep_2023}. While it would be difficult to train such a model on random data (as most datapoints would be trivial), the pre-filtration through our method makes the dataset more diverse and thus more likely to provide the model with useful examples.

Some other research directions involve certainly applying this method in a wider context, both considering other phase types, and other continuous systems. Systems such as Flow-Lenia \cite{plantec_flow-lenia_2023} where mass is conserved Fare particularly interesting because they require the design of a different type of phase (as 'dying' configurations don't exist). Another worthwhile direction would certainly be to include the initial condition in the varying parameter set, to see if phase transition region also increase the amount of interesting behaviors in that context.

In conclusion, this method opens avenues for parameter exploration which are mostly orthogonal to other methods such as genetic algorithms and reinforcement learning, and in combination they open exciting possibilities for automatic discovery of Artificial Life systems !

\section{Acknowledgements}
The authors would like to thank 
João Penedones, Eug\'ene Bergeron and Sanyam Jain for interesting discussions. Support
from the Blavatnik Family Foundation, the Latsis Foundation, the NCCR SwissMAP, and from an EPFL FSB Seed
Funding Grant are gratefully acknowledged.
\bibliographystyle{ACM-Reference-Format}
{\bibliography{sample-base}}

\appendix
\section{Code and simulations}
The full code can be found at \href{https://github.com/frotaur/TempGECCO}{https://github.com/frotaur/TempGECCO}. It includes the code to run the PTF algorithm on Lenia, as well as some particularly interesting parameter sets found with the method. The dynamics for these parameters can be simulated and viewed in real time with the codebase.

Alternatively, \href{https://vassi.life/teaching/alife/teaser}{https://vassi.life/teaching/alife/teaser} showcases a subset of the dynamics under the section 'Lenia'.
\end{document}